\documentstyle[12pt,aasms4]{article}
\received{}
\revised{}
\accepted{}

\cpright{}{}
\journalid{}{}
\articleid{}{}
\paperid{}

\cpright{}{}
\ccc{}

\lefthead{Kov\'acs and Jurcsik}
\righthead{On the light curve -- luminosity relation of RR Lyrae stars}

\begin{document}

\title{On the light curve -- luminosity relation of RR Lyrae stars}

\author{G\'eza Kov\'acs and Johanna Jurcsik}

\affil{Konkoly Observatory, P.O. ~Box ~67, H--1525,
Budapest, Hungary \\
electronic mail: kovacs, jurcsik(@buda.konkoly.hu)}

\authoraddr{Konkoly Observatory, P.O. ~Box ~67, H--1525,
Budapest, Hungary}

\begin{abstract}
We use a large sample of RRab stars in globular clusters and in 
the Sculptor dwarf galaxy to decipher the relation between the 
Fourier decomposition and the luminosity. For fixing the zero point 
of the relation, we use the latest Baade-Wesselink (BW) results.
It is shown that the most plausible representation of the absolute 
brightness (in V color) consists of a linear expression of the period 
and of the Fourier parameters $A_1$ and $\varphi_{31}$ (computed also 
from the V light curve). We derive an average 
$M_V = \alpha {\rm [Fe/H]} + \beta$ relation from all the available 
Fourier data. Our results exclude any values of $\alpha$ larger 
than 0.19, in agreement with most of the BW and evolutionary 
studies. We give age and reddened distance modulus estimations for 
the clusters entering in our analysis.
\end{abstract} 
%
%
\section{Introduction}

Determination of the luminosity of RR Lyrae stars is a high 
priority task in the study of these stars, because they play a key 
role as distance and age indicators in our Galaxy and in its close 
neighborhood. With the rapid progress of the massive photometry, 
the number of RR Lyrae stars with accurate light curves will soon 
be increased by one or two orders of magnitudes. To utilize these 
data in mapping the not too distant parts of the universe, we need 
accurate relations between the physical and the light curve 
parameters. Several previous attempts tried to tackle this question. 
In the simplest case a linear relation is sought for $M_V$ 
either with [Fe/H] (Clementini et al. 1995, and references therein)
or with [Fe/H] and the period $P$ (e.g. Nemec et al. 1994).
From the theoretical side we mention Simon and Clement (1993) who 
used nonlinear pulsation models of RRc stars to derive relations 
between the Fourier parameters and the luminosity. It is important 
to note that the often quoted dependence of $M_V$ on [Fe/H] is 
suggested by the evolutionary theories and is obtained through 
an {\it ensemble average} of many evolutionary tracks. The several 
empirical variants of this relation suffer from the ambiguity due 
to the limited number of stars used. In addition, because of the 
recently discovered tight dependence of [Fe/H] on $P$ and 
$\varphi_{31}$ (Jurcsik and Kov\'acs 1996, hereafter JK), the 
single parameter dependence of $M_V$ can clearly no longer be hold. 

In this paper we follow JK and derive a linear relation for the 
luminosity of the RRab stars. Up to the zero point, the 
result is {\it free} of any assumptions and relies solely on 
the empirical relation between {\it directly observed} quantities.  
%
%
\section{The data base}

For the purpose of the empirical determination of the relation 
between the luminosity and the shape of the light curve, 
we use the cluster data as compiled by JK. Here we omit the LMC 
clusters NGC 1835 and 2257 mostly because of the low number of 
the stars, but also because of their lower quality. Additional 
data come from the very recent OGLE observations of the Sculptor 
dwarf spheroidal galaxy and those of the galactic globular clusters 
Ruprecht 106 (Kaluzny et al. 1995a,b) and M5 (Reid 1996). We use 
(as in JK) Johnson V color throughout this paper. For the 
absolute calibration we apply the compilation  of the BW luminosities 
as given by Clementini et al. (1995) (their Table 21, column 5). 
All these data sets are listed in Table 1. Since we use only 
the best quality light curves without any peculiarity or sign 
of Blazhko behavior, the numbers shown in the table are lower 
than the ones given for the RRab stars in the respective 
publications. In the Fourier analysis we follow the same method 
as outlined in JK. Reddening corrections are applied only in the 
case of M4, because of its considerable differential extinction 
(Cacciari 1979; Liu and Janes 1990).
%
%
\section{The $(P,A_1,\varphi_{31}) \rightarrow M_V$ relation}

First we use only the globular cluster and the Sculptor data and do not 
rely on the BW luminosities. In finding the relation between the Fourier 
parameters and the luminosity we follow essentially the same method as 
in JK. 

Here we utilize on the fact that the distance moduli are the same for all 
stars in each cluster. In addition, it is assumed that the same thing 
is true also for the reddenings. Since the luminosity should follow a 
general relation, independently of the cluster considered, for a 
consistent representation of the apparent magnitudes we fit a certain 
number of Fourier parameters and $n-1$ constants, where $n$ is the 
number of clusters. Including the period and all the Fourier 
amplitudes and phases up to order 6, we search for the best linear 
relation representing the observed ({\it intensity averaged}) 
brightness. In this way we get an optimum set of 
{\it reddened relative distance moduli}, and a formula representing 
the relation between the light curve and the absolute magnitude. Stars 
with discrepant luminosities are left out of the fit. We do not discuss 
these stars in detail, we only remark that there might be several reasons 
for their peculiarity, e.g. crowding effects, inhomogeneous extinction, 
Blazhko behavior, etc. Although the fairly high noise level introduces 
some ambiguity in the selection process, we try to avoid it by keeping 
as many stars as possible, and reaching a situation when both the fitting accuracy and the regression coefficients seem to settle 
on a stationary value. We think that we reach this state by omitting 
5 stars from the globular clusters and 7 stars from Sculptor. Finally 
we end up with 177 stars. Since the fitting accuracy shows a leveling 
off when using more than 3 Fourier parameters (cf. Table 3), we settle 
on the following formula which fits the data with an accuracy of 
0.047 mag 
\begin{eqnarray}
M_V = const. - 1.398 P - 0.471 A_1 + 0.104 \varphi_{31}
                     \hskip 2mm .
\end{eqnarray}
To obtain an absolute calibration, we repeat the above process by 
including also the BW stars. We emphasize that we use the BW stars 
merely to {\it fix the zero point}. We find that the BW luminosities 
of the following stars do not confine to the basic trend of the 
clusters: WY Ant, UU Cet, DX Del, SS Leo, AV Peg, BB Pup, W Tuc, 
M5 V28, M92 V1, V3. After leaving out these stars, we get a set of 
198 stars which we refer hereafter as the {\it calibrating data set}. 
This yields the following expression 
\begin{eqnarray}
M_V = 1.221 - 1.396 P - 0.477 A_1 + 0.103 \varphi_{31}
                     \hskip 2mm ,
\end{eqnarray}
with the corresponding error formula
\begin{eqnarray}
\sigma^2_{M_V} = 0.2275\sigma^2_{A_1} + 0.0106\sigma^2_{\varphi_{31}} + 
                       \sum^4_{i,j=1} K_{ij} p_i p_j 
                       \hskip 2mm ,
\end{eqnarray}
where $p_1=1$, $p_2=P$, $p_3=A_1$, $p_4=\varphi_{31}$, and the 
correlation coefficients $K_{ij}$ are given in Table 2. In this 
formula we have omitted the completely negligible error of the 
period. We recall, that just as in JK, the Fourier parameters 
refer to a {\it sine} decomposition and that the phase should be 
chosen as the closest value to 5.1. 

In Fig. 1 we show the calculated vs. the observed absolute magnitudes 
for the best single and the above three parameter fits. The 'observed' 
absolute magnitudes are calculated by the use of the apparent 
brightnesses and the computed reddened distance moduli. It is clear 
that the inclusion of the Fourier parameters caused a visible 
improvement in the representation of the data compared with that 
of the traditional period -- luminosity relation. 

It is important to address the question of the significance of the 
three parameter relation over the two parameter one. In order to 
do so, we generate artificial data with the following formula
\begin{eqnarray}
M_V(i) = 1.736 - 1.242 P(i) - 0.659 A_1(i) + \xi(i)
                     \hskip 2mm ,
\end{eqnarray}
where $\xi(i)$ is a Gaussian random number with $\sigma_{\xi}=0.05$. 
The regular part of Eq. (4) corresponds to the best two parameter 
fit to the calibrating data set and the indices run through this set. 
For each realization of $\{\xi(i)\}$ we find the best $n$ parameter 
fit just like in the case of real data. Denoting the {\it unbiased} 
estimation of the standard deviation of the best $n$ parameter fit 
by $\sigma_n$, we ask what is the probability that the relative 
reduction of $\sigma_n$ (i.e. $1 - \sigma_n/\sigma_{n-1} \equiv \rho_n$) 
exceeds a certain limit when the three and two parameter fits are 
compared. After a large number of simulations, the distribution 
function shown in Fig. 2 is obtained. This can be used to estimate 
the significance of $\rho_3$ obtained in the case of the observed 
data of the calibrating set (Table 3). We see that $\rho_3$ is about 
3 times greater than the maximum reduction of the dispersion found 
in the random simulation. Therefore, we conclude that the three 
parameter formula (2) is statistically {\it highly significant} in 
the representation of the luminosity. In addition, as can be seen 
in Table 3, the fitting accuracy clearly levels off for the higher 
parameter regressions. This shows that the three parameter description 
not only necessary, but also sufficient for the representation of the observations.  

Finally we mention that because of the close interrelations among the 
amplitudes and phases (see JK), there exist compatible formulae which 
contain other Fourier components. Our Eq. (2) has not only the highest 
fitting accuracy, but also contains low order Fourier components, which 
helps to diminish observational errors.

%
%
\section{The ${\rm [Fe/H]} \rightarrow M_V$ relation}

In deriving an average ${\rm [Fe/H]} \rightarrow M_V$ relation we 
utilize the fact that with the aid of the formula of JK and Eq. (2) 
we are able to estimate both [Fe/H] and $M_V$ for any RRab star with 
reliable Fourier decomposition. Using all data of this paper and 
those of JK we get the result shown in Fig. 3. It is seen that the 
${\rm [Fe/H]} \rightarrow M_V$ relation suffers from a considerable 
{\it intrinsic scatter}, due to the extra dependence on the Fourier 
parameters. The straight line is a least squares fit and corresponds 
to the following expression
\begin{eqnarray}
M_V  = 0.19 {\rm [Fe/H]} + 1.04 \hskip 2mm .
\end{eqnarray}
This formula is in a nice agreement with the recent 
BW results as summarized by Clementini et al. (1995), and also 
with the evolutionary calculations (e.g. Lee 1990). 
Since Eq. (2) depends somewhat on the sample of stars used in its 
derivation, this dependence translates to Eq. (5). Our experiences 
show that for all reasonable samples the coefficients of Eq. (5) are 
always in the ranges of 0.19 -- 0.16 and 1.04 -- 0.99, respectively.
Furthermore, changing our greater values of [Fe/H] to the lower 
spectroscopic ones for M68, M92 and NGC 1841, we get only a slight  
decrease in the coefficients, namely they become 0.18 and 1.03 
respectively. 
 
Therefore, our results {\it undoubtedly exclude} large [Fe/H] 
coefficients as it is sometimes quoted in the literature 
(Buonanno et al. 1990; Longmore et al. 1990; Sandage 1993).  

%
%
\section{Distance moduli and ages}

With Eq. (2) it is easy to get reddened distance modulus which can 
be directly (i.e. {\it without} reddening correction) applied to 
estimate the age of a cluster if the apparent turn-off luminosity 
$V(TO)$ is known. The results are shown in Table 4. The distance 
moduli and the apparent turn-off luminosities are {\it reddened}, 
except for M4, where reddening was taken into account as mentioned in 
Section 2. Errors of the distance moduli correspond to the standard 
deviations of the distance moduli obtained for the stars in each 
cluster. Superscripts refer to the sources of $V(TO)$. Abundances are 
calculated according to JK. To estimate the ages we apply the formula 
of Straniero and Chieffi (1991, their Eq. (4)). We recall that using 
the reddened distance moduli and the observed (also reddened) $V(TO)$, 
the absolute magnitude of the turn-off, $M_V(TO)$ and therefore, the age 
can be directly estimated without the resort to an additional observable 
(i.e. the mean brightness difference between the horizontal branch and the turn-off point, the estimation of which introduces further ambiguities 
$-$ see Caputo and Degl'Innocenti 1995). The formal errors of the ages 
are between 1 and 2 Gyr. We remark that according to our formula, M5 
and Rup 106 exhibit a relatively large spread in the metallicity with $\sigma_{\rm [Fe/H]} \approx 0.2$. 

Because it is not the subject of this paper to discuss the 
ages of globular clusters, we just mention that there is 
an {\it age spread} among them, as it has also been stated by 
other studies applying low steepness in the    
${\rm [Fe/H]} \rightarrow M_V$ relation (e.g. Walker 1992b). 

%
%
\section{Conclusions}

On the basis of the latest observational material we have shown 
that the luminosity of RRab stars depends on three observables, 
namely on the period and on the Fourier parameters $A_1$ and 
$\varphi_{31}$ (Eq. (2)). The relative accuracy of the estimated 
absolute V magnitudes are better than $0.05$ mag, the standard 
deviation of the residuals of the calibrating data set. This is 
comparable to the corresponding value of the $P \rightarrow M_K$ 
regressions obtained from infrared photometry (Longmore et al. 
1990; Jones et al. 1992). We think that a considerable part of 
the scatter comes from the inhomogeneous reddening inside the 
clusters. In the absolute calibration there is still an error 
probably smaller than $\approx 0.1 - 0.15$ mag both in $M_V$ and 
in $M_K$, because they all rely on the Baade-Wesselink results. 
Our average ${\rm [Fe/H]} \rightarrow M_V$ relation (Eq. (5)) 
yields a similar dependence on [Fe/H] than those given by the recent 
Baade-Wesselink and evolutionary studies. We think that future, 
more accurate CCD measurements in clusters or galaxies with large 
[Fe/H] spread will drastically improve the correlation between 
the luminosity and the Fourier parameters. This will enable us 
to determine the luminosity of {\it any} RRab star, provided that 
we can measure its light curve with a reliable accuracy. 

%
%
\acknowledgements
We are very much indebted to Janusz Kaluzny for sending us the 
data on Sculptor and on Ruprecht 106, which were recently obtained 
within the framework of the OGLE project. Grateful acknowledgements 
are also due to Neill Reid for his excellent data on M5. A part of 
this work was completed during G.K.'s stay at the Copernicus 
Astronomical Center in Warsaw. The supports of the Polish and the 
Hungarian Academy of Sciences and of OTKA grant T$-014183$ are 
acknowledged.

%
%
\newpage
\begin{table}
\caption{Data sets and their sources \label{Table 1}}
\begin{flushleft}
\begin{tabular}{ccc}
\tableline
\tableline
Name & N & Source \cr
\tableline
Gal. clust. & 71 & JK, Kaluzny et al. (1995b), Reid (1996) \cr
LMC         & 25 & JK \cr
Sculptor    & 93 & Kaluzny et al. (1995a) \cr
BW          & 31 & Clementini et al. (1995) \cr
\tableline
\tableline
\end{tabular}
\end{flushleft}
\end{table}

\begin{table}
\caption{Correlation coefficients $K_{ij}(=K_{ji})$ in Eq. (3) 
\label{Table 2}}
\begin{flushleft}
\begin{tabular}{ccrccr}
\tableline
\tableline
i & j &\multicolumn{1}{c} {$K_{ij}$} & i & j &\multicolumn{1}{c} {$K_{ij}$} \cr
\tableline
1 & 1 & $ 0.0101309 $ & 2 & 3 & $ 0.0020546 $ \cr
1 & 2 & $-0.0018854 $ & 2 & 4 & $-0.0001767 $ \cr
1 & 3 & $-0.0056066 $ & 3 & 3 & $ 0.0048972 $ \cr
1 & 4 & $-0.0014288 $ & 3 & 4 & $ 0.0005610 $ \cr
2 & 2 & $ 0.0036403 $ & 4 & 4 & $ 0.0002674 $ \cr
\tableline
\tableline
\end{tabular}
\end{flushleft}
\end{table}

\begin{table}
\caption{Variation of the standard deviation $\sigma_n$ and its 
relative reduction $\rho_n$ as a function of the number of parameters 
fitted to the calibrating data set
\label{Table 3}}
\begin{flushleft}
\begin{tabular}{clcr}
\tableline
\tableline
n & Params. & $\sigma_n$ & \multicolumn{1}{c}{$\rho_n$} \cr
\tableline
1 & $P$                                           & 0.0597 & $ 0.245$ \cr
2 & $P$,\ $A_1$                                   & 0.0499 & $ 0.164$ \cr
3 & $P$,\ $A_1$,\ $\varphi_{31}$                  & 0.0471 & $ 0.056$ \cr
4 & $P$,\ $A_1$,\ $\varphi_{31}$,\ $\varphi_{51}$ & 0.0470 & $ 0.002$ \cr
5 & $P$,\ $A_1$,\ $\varphi_{31}$,\ ...            & 0.0465 & $ 0.011$ \cr
6 & $P$,\ $A_1$,\ $\varphi_{31}$,\ ...            & 0.0465 & $ 0.000$ \cr
7 & $P$,\ $A_1$,\ $\varphi_{31}$,\ ...            & 0.0466 & $-0.002$ \cr
8 & $P$,\ $A_1$,\ $\varphi_{31}$,\ ...            & 0.0467 & $-0.002$ \cr
\tableline
\tableline
\end{tabular}
\end{flushleft}
\end{table}

\begin{table}
\caption{Reddened distance moduli and ages (in [Gyr]) of some 
globular clusters and the dwarf galaxy Sculptor \label{Table 4}}
\begin{flushleft}
\begin{tabular}{lccccc}
\tableline
\tableline
Cluster & Dist. mod. & $V(TO)$ & $M_V(TO)$ & [Fe/H] & Age \cr
\tableline
M4 \tablenotemark{*}    & 11.17$\pm$0.02 & 15.60\tablenotemark{[1]}  & 4.43 & $-1.03$ & 18.8 \cr
M5     & 14.24$\pm$0.05 & 18.60\tablenotemark{[2]} & 4.36 & $-1.14$ & 18.1 \cr
M68    & 14.91$\pm$0.04 & 19.05\tablenotemark{[3]} & 4.14 & $-1.76$ & 17.6 \cr
M92    & 14.44$\pm$0.04 & 18.70\tablenotemark{[2]} & 4.26 & $-1.98$ & 21.2 \cr
M107   & 14.77$\pm$0.06 & 19.20\tablenotemark{[4]} & 4.43 & $-0.75$ & 17.0 \cr
N3201  & 13.95$\pm$0.05 & 18.25\tablenotemark{[5]} & 4.30 & $-1.33$ & 18.1 \cr
Ru106  & 16.99$\pm$0.03 & 21.05\tablenotemark{[6]} & 4.06 & $-1.51$ & 15.0 \cr
N1466  & 18.57$\pm$0.03 &  $-$  & $-$  & $-1.54$ &  $-$ \cr
N1841  & 18.62$\pm$0.03 & 22.60\tablenotemark{[7]} & 3.98 & $-1.73$ & 14.9 \cr
Retic. & 18.28$\pm$0.03 & 22.50\tablenotemark{[8]}& 4.22 & $-1.45$ & 17.3 \cr
Sculp. & 19.36$\pm$0.05 &  $-$  & $-$  & $-1.49$ &  $-$ \cr
\tableline
\tableline

\end{tabular}
\end{flushleft}
\tablenotetext{*}{Dereddened distance modulus and $V(TO)$}
\tablenotetext{[1]}
{Kanatas et al. 1995} 
\tablenotetext{[2]}{Buonnano et al. 1989}
\tablenotetext{[3]}{Walker 1994}
\tablenotetext{[4]}{Ferraro \& Piotto 1992}
\tablenotetext{[5]}{Brewer et al. 1993}
\tablenotetext{[6]}{Buonnano et al. 1993}
\tablenotetext{[7]}{Walker 1990}
\tablenotetext{[8]}{Walker 1992a}
\end{table}

\newpage

\centerline{\bf{FIGURE CAPTIONS}}

\figcaption[mv1.eps]{Calculated vs. observed $M_V$ relations. 
{\it Top:} three parameter fit (Eq. (2)); 
{\it bottom:} single parameter fit. The calibrating data set is plotted. 
Open circles denote the BW stars. The $45^{\circ}$ lines are shown for 
guidance. 
\label {Fig. 1}}

\figcaption[mv2.eps]{Probability distribution of the relative 
reduction of the standard deviation for artificially generated data 
(Eq. (4), see text for details). The three parameter models are tested 
against a two parameter one. 
\label {Fig. 2}}

\figcaption[mv3.eps]{The relation between ${\rm [Fe/H]}$ and $M_V$ 
for 278 stars with reliable Fourier decompositions. ${\rm [Fe/H]}$ 
and $M_V$ are computed from Eq. (3) of JK and Eq. (2) of this paper. 
The linear regression (Eq. (5)) is shown by solid line.
\label {Fig. 3}}

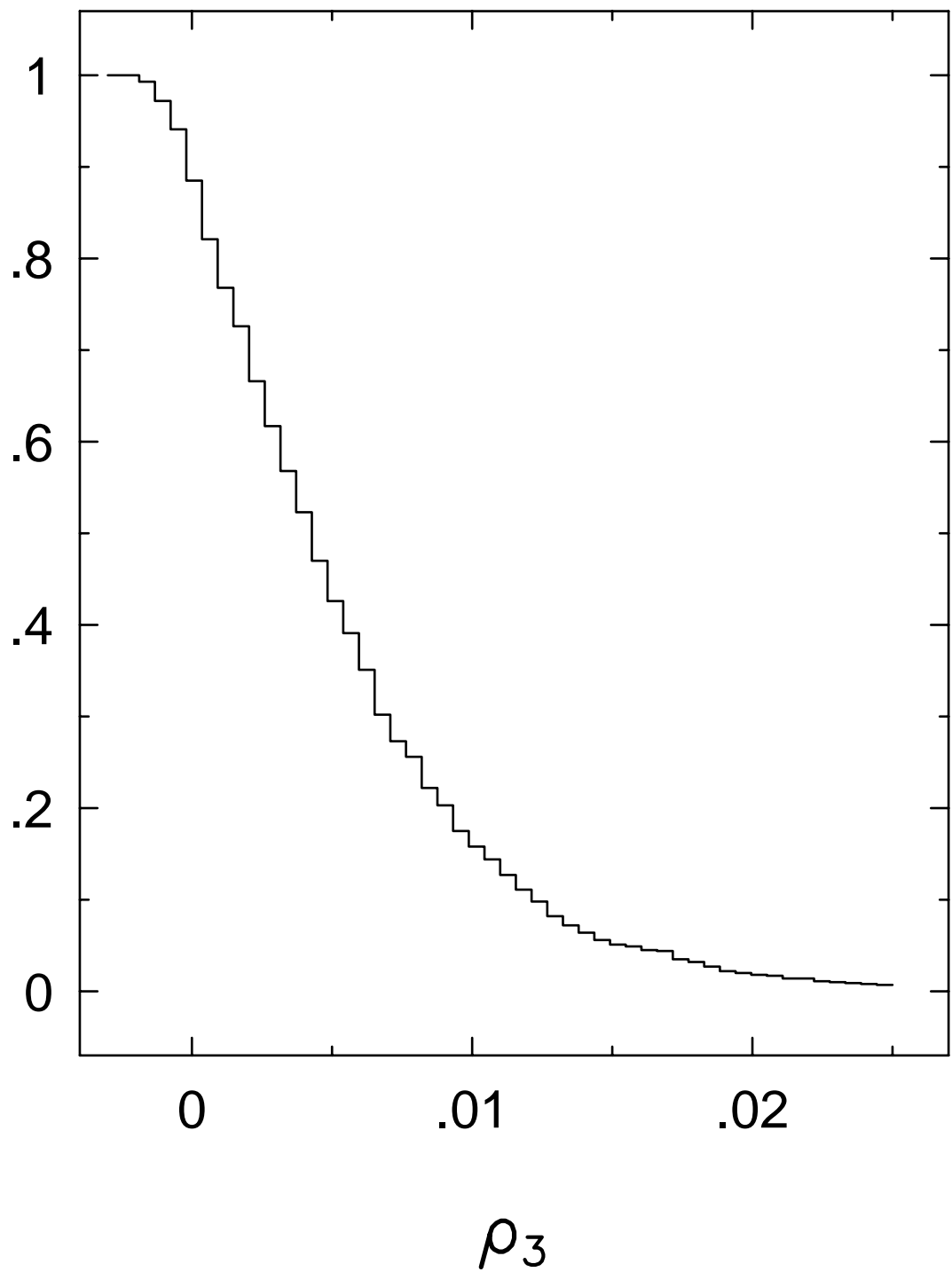
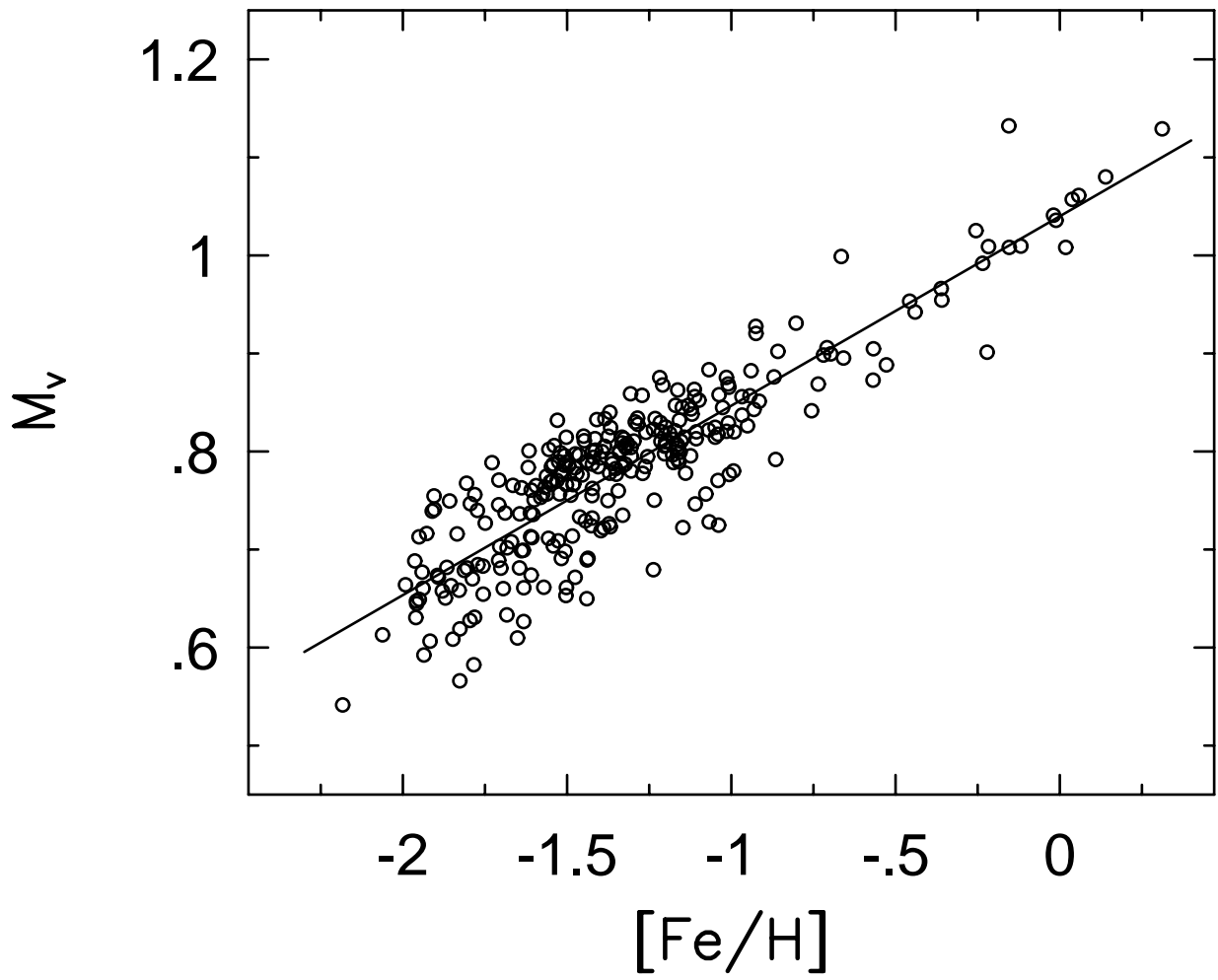
\end{document}